\documentclass[twocolumn,pra,showpacs,preprintnumbers,amsmath,amssymb]{revtex4}
\usepackage{graphicx}
\usepackage{dcolumn}
\usepackage{bm}
\usepackage{empheq}
\usepackage{color}
\newcommand{\sn}{\textrm{sn}}
\newcommand{\cn}{\textrm{cn}}
\newcommand{\dn}{\textrm{dn}}
\newcommand{\am}{\textrm{am}}
\newcommand{\sd}{\textrm{sd}}
\newcommand{\sign}{\textrm{sign}}

\definecolor{green1}{rgb}{0.,0.9,0.}
\newcommand{\green}{\textcolor{black}}

\begin{document}
\title{Doubly periodic solutions of the focusing nonlinear Schr\"odinger equation: recurrence, period doubling and amplification outside the conventional modulation instability band}
\author{Matteo Conforti,$^{1*}$ Arnaud Mussot,$^{1}$ Alexandre Kudlinski,$^{1}$ Stefano Trillo$^{2}$ and Nail Akhmediev$^{3}$}
\affiliation{%
$^{1}$University of Lille, CNRS, UMR 8523—PhLAM—Physique des Lasers Atomes et Mol\'ecules, Lille, France\\
$^{2}$Department of Engineering, University of Ferrara, Ferrara, Italy \\
$^{3}$Optical Sciences Group, Department of Theoretical Physics, Research School of Physics, The Australian National University, Canberra, ACT 2600, Australia
}%

\date{\today}

\begin{abstract}
Solitons on a finite a background, also called breathers, are solutions of the focusing nonlinear Schr\"odinger equation, which play a pivotal role in the description of rogue waves and modulation instability. 
The breather family includes Akhmediev breathers (AB), Kuznetsov-Ma (KM), and Peregrine solitons (PS), which have been successfully exploited to describe several physical effects. These families of  solutions are actually only particular cases of a more general three-parameter class of solutions originally derived by Akhmediev, Eleonskii and Kulagin [Theor. Math. Phys. {\bf 72}, 809--818 (1987)]. Having more parameters to vary, this significantly wider family has a potential to describe many more physical effects of practical interest than its subsets mentioned above. The complexity of this class of solutions prevented researchers to study them deeply.
In this paper, we overcome this difficulty and report several new effects that follow from more detailed analysis. Namely, we present the doubly periodic solutions and their Fourier expansions.
In particular, we outline some striking properties of these solutions. Among the new effects, we mention (a) regular and shifted recurrence, (b) period doubling and (c) amplification of small periodic perturbations with frequencies outside the conventional modulation instability gain band.
\end{abstract}


\maketitle

\section{Introduction}

The nonlinear Schr\"odinger equation (NLSE) is one of the paradigms of modern nonlinear science. It describes the evolution of narrow-band envelopes under the combined action of weak dispersion and non-linearity, and naturally appears in different branches of physics such as optics \cite{Agrawal}, hydrodynamics \cite{Zak1968}, plasma \cite{Hasegawa} and cold atoms \cite{Kevrekidis}.

In the focusing regime, the continuous wave (CW) solution of NLSE is unstable with respect to modulation instability (MI) \cite{Bespalov1966,Benjamin1967}, which entails the exponential growth of low-frequency perturbations. Modulation instability is the most basic and widespread nonlinear phenomenon, which has been studied since the 60s in such a diverse disciplines as hydrodynamics \cite{Benjamin1967} or nonlinear optics \cite{Bespalov1966}. 
While the initial development of MI that arises from the linear stability analysis is well understood, the nonlinear stage of MI is an extremely hot and active research topic. 
In the fully nonlinear regime, MI can give rise to Fermi-Pasta-Ulam recurrences \cite{AA42,TW1991,Van2001,Bend2015,Kim2016,Mus2018,Pie2018,Gri2018,Nav2019a}, modulated cnoidal waves \cite{Kar1967,El1993,Bio2016,Bio2018,Kra2019a}, deterministic formation of breathers \cite{Kib2010,Kib2012,Zak2013,Gel2014,Kib2015,Nav2019b}, whose appearance can be triggered by the shape of the initial perturbation of the background \cite{Con2018,Tri2019},
as well as tubulent states mediated by statistical appearance of breathers \cite{Soto2016,Aga2016,Kra2019b}. Moreover, MI has been identified as one of the possible generating mechanisms of rogue waves \cite{Dud2014,Bar2014,Bar2015}. The latter can be described analytically as breathers, or solitons on finite background. \\
\indent All the solutions that describe such regimes can be presented in analytic form due to the integrability of the NLSE and the inverse scattering transform (IST) being the powerful tool for its analysis.
There are  different formulations of IST depending on specific boundary conditions of the problem to be solved. For example, problems related to MI require periodic boundary conditions. These set of problems is more complicated than the one with zeros at infinity. Thus, special finite-band integration theory has been developed \cite{Bel1994,Mat2008,Kam1997,Ma1981}. This technique permits one to write the solutions of the NLSE with periodic boundary conditions explicitly as ratios of Riemann theta functions \cite{Its1976,Osb2002,Gri2018b,Tra1984,Tra1988}.
However, these formal solutions contain an infinite number of free parameters and their practical use remains questionable. Indeed, extracting physically relevant solutions from these general ones is not an easy task. Practical solutions are mostly relying on direct methods with the most popular one being the Hirota method \cite{Nak1980}.

An alternative way to proceed is constructing more complex solutions from simple ones rather than trying to extract simple solutions from the most general one.
One of these techniques is Darboux transform that allows one to start with seeding solution in the form of a plane wave and build periodic solutions with elaborate evolution \cite{Akhbook,Akh1988}. Using the plane wave as a seed solution only allows one to find some classes of solutions that can be considered as higher-order MI. In order to expand this class of solutions to more general ones, other types of seed solutions must be used.

The basic family of solutions of the NLSE that is periodic both in space and time can be found using a special ansatz suggested in \cite{Akh1986,Akh1987}. 
This technique allows one to reduce the NLSE which is an infinite dimensional Hamiltonian system to a finite dimensional one that can be then solved analytically \cite{Akhbook,Akh1986,Akh1987,Mihalache1993}. The result is the family of first-order solutions that have arbitrary amplitude and arbitrary periods in time and space \cite{Akh1987}. Well known solutions such as Peregrine soliton (PS) \cite{Per1983}, Kuznetsov-Ma (KM) soliton \cite{Kuz1977,Ma1977} or Akhmediev breather (AB) \cite{Akh1986} and the two-parameter family of doubly periodic solutions of A and B-types \cite{Akh1986,PRA2017,Kim2016} are special cases of this significantly wider family. Thus, we can talk about this three parameter family as the most general \green{known} family of doubly periodic solutions.

In this paper, we report a detailed study of the properties of the first-order solutions, which include all above listed subsets and provide an analytic description for several new intriguing nonlinear phenomena.
These include: (1) FPU recurrence as a natural consequence of longitudinal periodicity, (2) period doubling during the MI process of growing periodic perturbation and (3) the MI growth outside of the conventional MI band. For the first time, we found the Fourier coefficients of the first order solutions in analytic form. The latter is an essential part of the present work that made it possible the expansion of the number of physical applications of the doubly periodic solutions.

The paper is organised as follows. In Sec. II, we give the exact form of the family of solutions with three free parameters. Moreover, we present the expressions of the Fourier coefficients and analyse the relevant physical properties such as the amplitude, the period and the wavenumber of the solutions as a function of these three governing parameters. In Sec. III we predict, based on these solutions, the new counter-intuitive phenomena such as the growth of linearly stable perturbations and period doubling within the MI. In Sec. IV we summarise our results.

\section{First-order doubly periodic solutions}

We start from the NLSE written in the following normalised form, which is the standard nonlinear fiber optics notation:
\begin{equation} \label{nlse}
i\psi_z+\frac{1}{2}\psi_{tt}+|\psi|^2\psi=0.
\end{equation}

The first-order solutions have the following property: at each propagation step $z$, the real and imaginary parts of the field $\psi(t,z)$ are linearly related \cite{Akh1986}. This implies the following form of the solution \cite{Akh1987}:
\begin{equation}\label{ansatz}
\boxed{\psi(t,z)=[Q(t,z)+i\delta(z)]e^{i\phi(z)}},
\end{equation}
where $Q$, $\delta$ and $\phi$ are real functions. 
For every value of $z$, Eq.(\ref{ansatz}) represents a straight line in the complex plane ($Re\psi$,$Im\psi$).
The ansatz (\ref{ansatz}) permits to reduce the NLSE, which is an infinite-dimensional Hamiltonian system, to a finite number of dimensions. The unknown functions $Q$, $\delta$ and $\phi$ are calculated through the solution of three nonlinear ordinary differential equations. In the following, we present only the final forms of the solutions, the method being thoroughly discussed in \cite{Akh1987,Akhbook,Mihalache1993}.

The first-order solutions are in general doubly periodic, i.e. periodic both in time and space  and depend on three real parameters. We classify the solutions in two types the same way as in \cite{Akh1986} (see Fig. 1 of \cite{Akh1986}). 
The type A solutions describe shifted recurrence, where local maxima in time appear with a shift of half temporal period  after a propagation distance corresponding to a half of spatial period. The type B solutions describe the recurrence, where the local maxima in time appear at the same temporal position after a propagation distance corresponding to one spatial period. 
 The separatrix between these two families, known as Akhmediev breather, is a heteroclinic orbit (periodic in time, non periodic is $z$), connecting the two CW solutions of the same amplitude but different phases.The convention taken in the recent work \cite{PRA2017} is the same as here. However, the notation used in \cite{Kim2016} is reversed.

The family of solutions is controlled by three parameters $\alpha_{1}$, $\alpha_{2}$ and $\alpha_{3}$ which are the three roots of a fourth order polynomial, with the fourth one being zero \cite{Akh1987}. They can either be all real, ordered in such a way that $\alpha_3>\alpha_2>\alpha_1>0$, or one real and two complex conjugates: $\alpha_3>0$ and $\alpha_1=\alpha_2^*=\rho+i\eta$. Division of the the roots $\alpha_i$ by some positive number $C$ is equivalent to transition from the solution $\psi(t,z)$ corresponding to the roots $\alpha_i$ to a different solution $\psi'(t,z)$ corresponding to the roots $\alpha_i/C$. The two solution are connected by the transformation
\begin{equation}\label{trans}
\psi(t,z)=C\psi'(Ct,C^2z).
\end{equation}
As a constant $C$, we can choose, for example, the value of one of the roots and seek a two-parameter family of solutions. The third parameter can be reintroduced by means of the transformation (\ref{trans}), thus adjusting the amplitude. 
In the following, we consider $\alpha_3=1$, which physically means fixing the initial CW component of the solution (strictly for the A-type solutions and approximately for the B-type). The two parameters $\alpha_1$ and $\alpha_2$ allow us to tune independently the spatial and temporal periods of the solutions. In practice, this means that if the period and amplitude of the initial condition are known, we can predict the oscillating pattern of the resulting evolution.

\subsection{The B-type solutions}
As explained above, here, we stick to the definition of doubly periodic solutions introduced in \cite{Akh1986}. Namely, A-type solutions are those located inside the separatrix corresponding to the Akhmediev breather while B-type solutions are located outside of the separatrix.
We start with the B-type solutions. They
depend on the three real parameters $\alpha_3>\alpha_2>\alpha_1>0$.

For the  function $\delta(z)$, we have the following expression:
\begin{equation}
\boxed{\delta(z)=\frac{\sqrt{\alpha_1\alpha_3}\sn(\mu z,k)}{\sqrt{\alpha_3-\alpha_1\cn^2(\mu z,k)}}},
\end{equation}
where the modulus $m$ of the Jacobian elliptic functions \cite{Byrd,Abramowitz} is $m=k^2=\displaystyle\frac{\alpha_1(\alpha_3-\alpha_2)}{\alpha_2(\alpha_3-\alpha_1)}$ and $\mu=2\sqrt{\alpha_2(\alpha_3-\alpha_1)}$. It is important to note that $0<\delta^2<\alpha_1$ by construction.

For the  function $\phi(z)$, we have the following expression:
\begin{equation}
\boxed{\phi(z)=(\alpha_1+\alpha_2-\alpha_3)z+\frac{2\alpha_3}{\mu}\Pi(\am(\mu z,k),n,k)},
\end{equation}
where $n=\displaystyle\frac{\alpha_1}{\alpha_1-\alpha_3}$ and $\Pi(\am(\mu z,k),n,k)$ is the incomplete elliptic integral of the third kind, with the argument $\am(u,k)$ being the amplitude function \cite{Byrd,Abramowitz}.

For the function of two variables $Q(t,z)$, we have the following expression:
\begin{equation}\label{Qtype1}
\boxed{Q(t,z)=\frac{Q_D(Q_A-Q_C)+Q_A(Q_C-Q_D)\sn^2(p\,t,k_q)}{(Q_A-Q_C)+(Q_C-Q_D)\sn^2(p\,t,k_q)}} ,
\end{equation}
where the elliptic modulus is $m_q=k_q^2=\displaystyle\frac{\alpha_2-\alpha_1}{\alpha_3-\alpha_1}$ and $p=\sqrt{\alpha_3-\alpha_1}$. Importantly, $Q_C<Q<Q_D$ by construction.
The $z$-dependent functions $Q_A(z)>Q_B(z)>Q_C(z)>Q_D(z)$ are defined by the following expressions:
\begin{align}
\label{QA} Q_A&=\phantom{-} s\sqrt{\alpha_1-y}+\sqrt{\alpha_2-y}+\sqrt{\alpha_3-y},\\ 
Q_B&=-s\sqrt{\alpha_1-y}-\sqrt{\alpha_2-y}+\sqrt{\alpha_3-y},\\
Q_C&=-s\sqrt{\alpha_1-y}+\sqrt{\alpha_2-y}-\sqrt{\alpha_3-y},\\
\label{QD} Q_D&=\phantom{-}s\sqrt{\alpha_1-y}-\sqrt{\alpha_2-y}-\sqrt{\alpha_3-y},
\end{align}
where $y(z)=\delta^2(z)$, and $s=s(z)=\sign(\delta_z)=\sign(\cn(\mu z,k))$. The expressions in Eqs. (\ref{QA})-(\ref{QD}) differ from the original ones given in \cite{Akh1987} by the sign function $s$ in the first terms. This amendment removes the discontinuities of the $z$-derivative of the field $\psi$, which were present in the original formulation.

The period along $z$ ($=L$) and along $t$ ($=T$) can be calculated as follows:
\begin{equation}\label{periodA}
L=\frac{4K(k)}{\mu},\;\;T=\frac{2K(k_q)}{p},
\end{equation}
where $K(k)$ is the complete elliptic integral of the first kind \cite{Byrd}.

\begin{figure}[ht]
	\centering
		\includegraphics[width=\columnwidth]{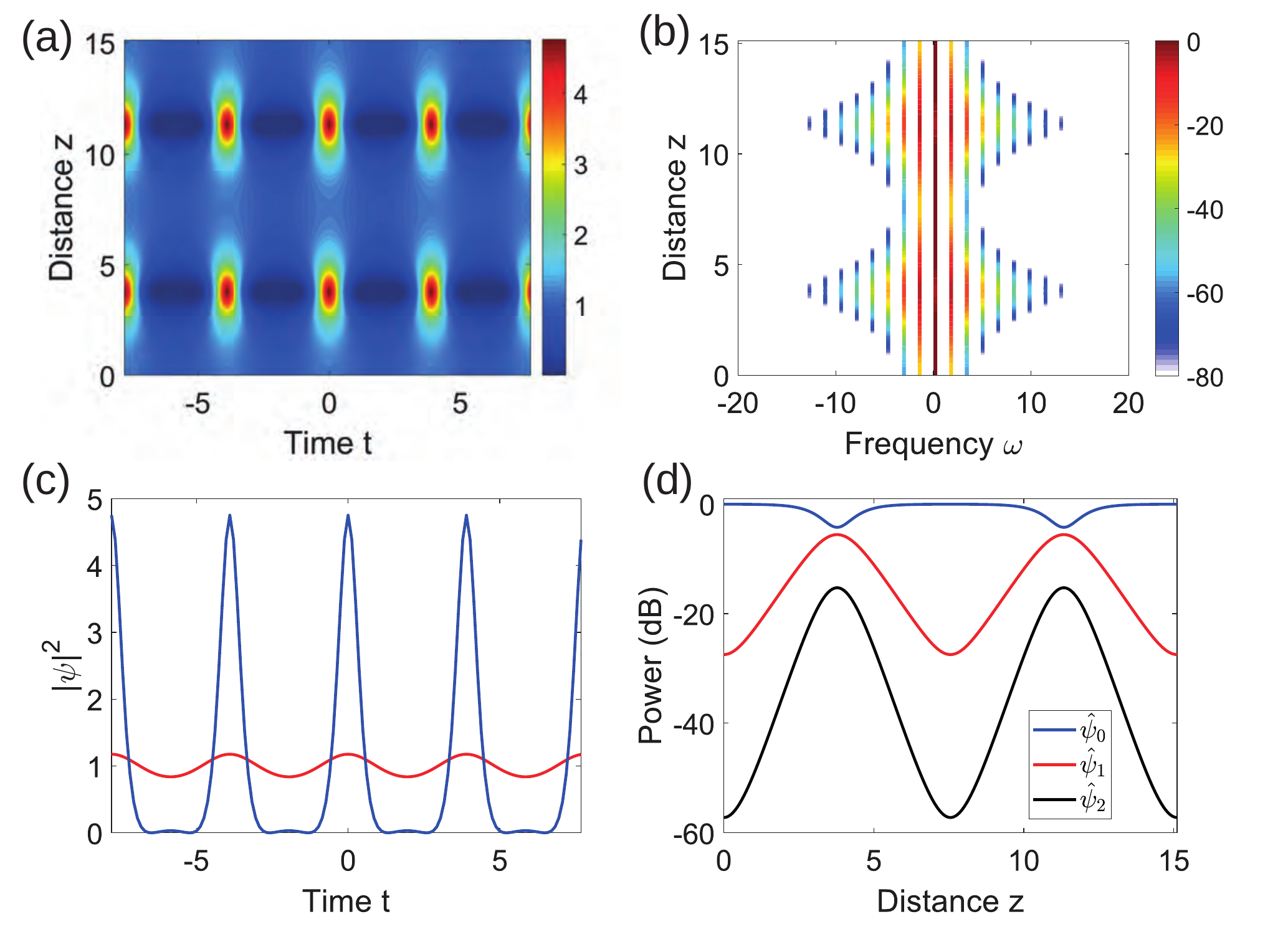}	
	\caption{An example of the B-type double-periodic solution. (a) False colour plot of intensity $|\psi(z,t)|^2$. Two longitudinal periods are shown. (b) Evolution of the spectrum. (c) Intensity profile $|\psi(t)|^2$ at $z=0$ (red curve, minimum of modulation) and at $z=L/2=3.78$ (blue curve, maximal modulation). (d) Evolution of the first three Fourier components in logarithmic scale ($20\log_{10}|\hat\psi_k(z)|$). Parameters: $\alpha_1=0.3$, $\alpha_2=0.4$, $\alpha_3=1$, giving the temporal period $T=3.9$ (modulation frequency $\omega=2\pi/T=1.61$) and the spatial period $L=7.55$.} 
	\label{fig1}
\end{figure}

A typical example of the B-type solution is shown in Fig. \ref{fig1}. Parameters of the solution are given in the figure caption. Figure \ref{fig1}(a) shows the spatio-temporal evolution of the \green{intensity} $|\psi|^2$, and Fig. \ref{fig1}(c) shows the intensity profile at the input ($z=0$, red curve) and at the point of the maximum temporal pulse compression ($z=L/2\approx3.8$, blue curve).  With the given set of parameters, the solution is similar to the evolution of the separatrix (AB) \green{up to $z\approx 7.5$}. It describes the amplification of a small but finite periodic modulation on top of a strong CW towards the generation of a periodic train of pulses. Just as the separatrix, after the maximum compression, the field returns back to its initial profile. However, in contrast to the separatrix, this happens at finite length $z=L=7.55$. In spectral domain, the energy spreads from the central component to sidebands with the following return back to the central one. We have, thus, an analytic description of the FPU recurrence phenomenon except that the initial spectrum  is not completely concentrated in a single component. An important point is that the presence of three independent parameters of the family allows us to control independently periods in $x$ and $t$ and the amplitude of the input. This is discussed, in more detail, below.

\subsection{Fourier spectrum of the B-type solution}
From practical point of view, important parameters of periodic solutions are their spectral components. Remarkably, for the above solutions, the calculations can be done analytically.
Being periodic in $t$ variable, the $Q$ function in Eq. (\ref{Qtype1}) can be expanded in Fourier series as follows:
\green{
\begin{align}
\nonumber Q(t,z)&=\sum_{\ell=-\infty}^{+\infty}\hat Q_\ell(z)e^{ i\frac{2\pi\ell }{T}t}\\
&=\hat Q_0(z) +2\sum_{\ell=1}^{+\infty}\hat Q_\ell(z)\cos\left(\frac{2\pi\ell }{T}t\right),
\end{align}
}
where the $z$-dependent Fourier coefficients \cite{Lang} are:
\begin{empheq}[box=\fbox]{align}
\label{fourierA0} \hat Q_0(z)&=Q_D+(Q_D-Q_A)\frac{\Pi(n,k_q)}{K(k_q)},\\
\label{fourierA} \hat Q_\ell(z)&=(Q_D-Q_A)\frac{\pi \lambda}{2K(k_q)}\frac{\sinh(2\ell w)}{\sinh(2\ell w_0)}.
\end{empheq}
Here,
$n=(Q_D-Q_C)/(Q_A-Q_C),$ $$\displaystyle\lambda=\sqrt{\frac{n}{(n-1)(m_q-n)}},$$ $$w=\displaystyle\frac{\pi [K(k'_q)-v_0]}{2K(k_q)}, ~~~w_0=\displaystyle\frac{\pi K(k'_q)}{2K(k_q)},$$ $$v_0=F(\sin^{-1}(1-n)^{-1/2},k'_q), ~~~ k'^2_q=1-k^2_q$$ and $F(\varphi,k)$ is the incomplete elliptic integral of the first kind \cite{Byrd,Abramowitz}.
The evolution of the Fourier coefficients for the total field $\psi(x,t)$ is simply given by:
\begin{align}
\hat\psi_0(z)&=[\hat Q_0(z)+i\delta(z)]e^{i\phi(z)},\\
\hat\psi_\ell(z)&=\hat Q_\ell(z)e^{i\phi(z)}.
\end{align}

The evolution of the spectrum of the B-type solution for the same set of parameters as in Fig. \ref{fig1}(a) is shown in Fig. \ref{fig1}(b). The Fourier spectrum is symmetric with respect to 
$\omega=0$. For the choice of parameters in Fig. \ref{fig1}, the initial power of the CW component is approximately 1 (0 dB). The total range of changes shown here is 80 dB. Five Fourier components of the initial profile are located within this range. Namely,  $|\hat\psi_0(0)|\approx 0$ dB, $|\hat\psi_{\pm1}(0)|=-27.5$ dB, $|\hat\psi_{\pm2}(0)|=-57.2$ dB.  At the point of maximal pulse compression, the solution becomes a comb of 17 spectral lines that has a triangular shape. Indeed, the power of the harmonics decays as a geometrical progression with the order, as can be seen from Eq.(\ref{fourierA}). 
As the solution is $z$-periodic, it describes an energy cascade towards higher harmonics, which is repeatedly reversed back to the original spectrum. The detailed evolution of the first three spectral components is shown in Fig. \ref{fig2}(c). From this figure and from Eq.(\ref{fourierA}), it can be seen that the sidebands never vanish, i.e. the solution never turns to CW with $|\psi|=$ constant. This only happens for the special choice of parameters when the doubly periodic solution approaches the separatrix.

\subsection{Major characteristics of the B-type solutions}
As mentioned, the family of the B-type solutions has 3 variable parameters. Only two of them can be conveniently shown on a plane. 
Thus, we keep the scaling parameter $\alpha_3$ fixed. It relates the amplitude and the two periods along the $z$ and $t$ axes. Thus, it can be used in practical calculations for adjusting any of these parameters to the actual values obtained in experiments. Here, we present the two periods (or the corresponding frequency and wavenumber) and the maximal amplitude for fixed $\alpha_3=1$.

\begin{figure}[ht]
	\centering
		\includegraphics[width=\columnwidth]{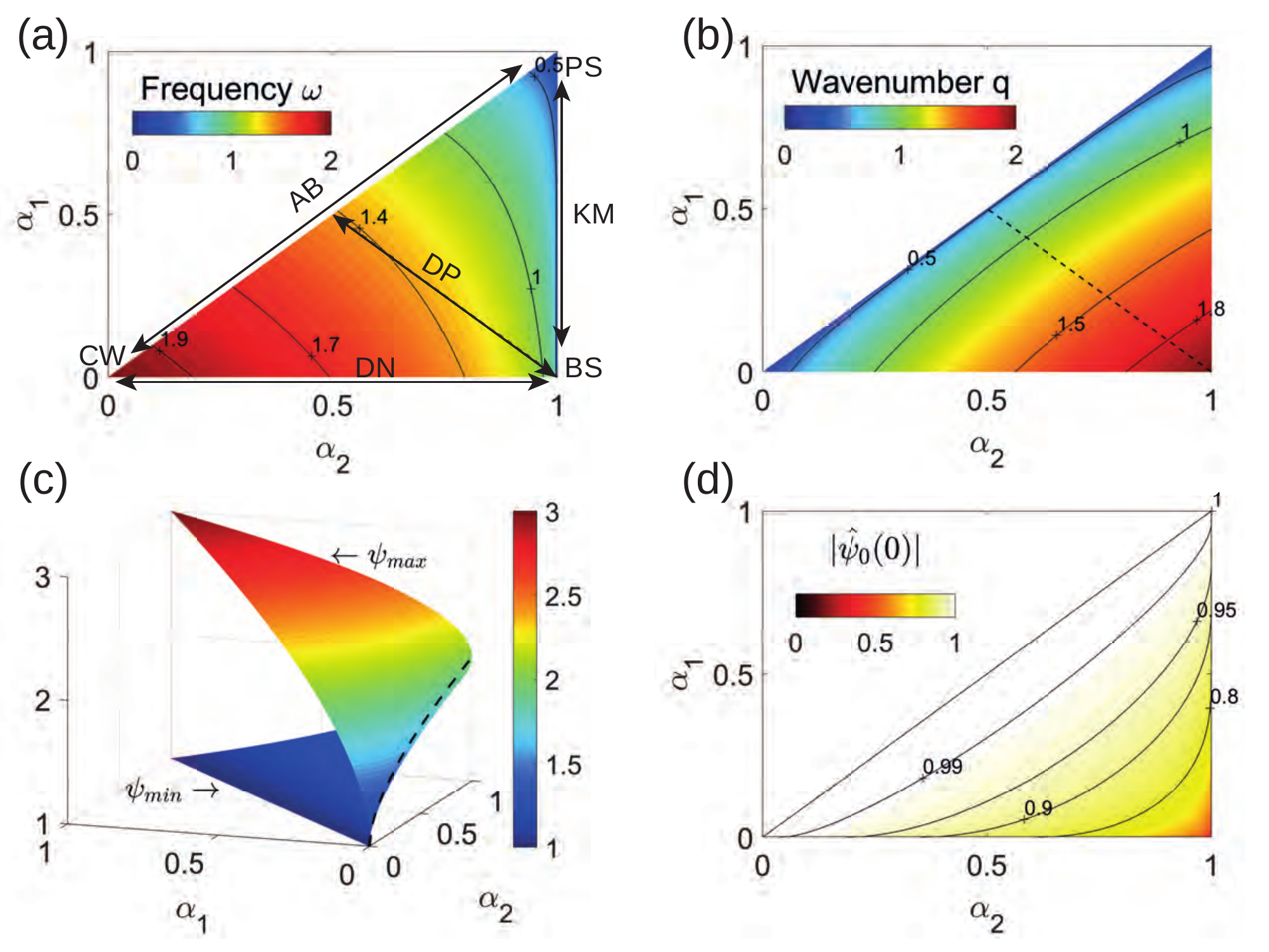}	
\caption{Major characteristics of the B-type doubly periodic solutions on the \green{plane} ($\alpha_1$,$\alpha_2$) when $\alpha_3=1$. 
(a) False color plot of frequency $\omega=2\pi/T$. (b) False color plot of wavenumber $q=2\pi/L$. The thin curves in (a) and (b) show the lines of equal frequency or wavenumber. (c) Maximal and minimal amplitudes of $|\psi(t=0,z)|$. (d) Absolute value of the CW component in $\psi(x,t)$ at $z=0$. Special cases in (a): PS - Peregrine solution, CW - the family of continuous wave solutions, DN - the family of DNoidal wave solutions, AB - the family of Akhmediev breathers, KM - the family of Kuznetsov-Ma solitons, BS - bright soliton on zero background. DP - the two parameter family of doubly periodic solutions.}
	\label{fig2}
\end{figure}

We recall that parameters $\alpha_1$ and $\alpha_2$ vary in the intervals of values from $0$ to \green{$\alpha_3=1$}. \green{Moreover, we have assumed that the roots are ordered in such a way that $\alpha_2>\alpha_1$. } 
 Thus, it is sufficient to consider these parameters within the triangular area shown in colour in Figs.\ref{fig2}(a) and \ref{fig2}(b). Figure \ref{fig2}(a) shows the frequency $\omega=2\pi/T$ while Figure \ref{fig2}(b) shows the wavenumber $q=2\pi/L$ both calculated using Eq.(\ref{periodA}). The crucial observation is that the range of admitted frequencies for the B-type solutions coincides with the modulational instability band $0<\omega<2$. Then, it is not surprising that the amplification of small periodic perturbations as a major feature of modulation instability remains valid at small deviations from the separatrix as we have seen in the previous Subsection.

However, the family of the B-type solutions includes much wider set of solutions.
When $\alpha_1=\alpha_2$, i.e. on the diagonal line in Figs.\ref{fig2}(a,b), the solution becomes homoclinic or a separatrix, known as Akhmediev breathers.
The members of this family of solutions have finite periods in $t$ and infinite periods in $z$. To be specific, period $L$ in this limit tends to infinity. 
At the point $\alpha_1=\alpha_2=1$, the solution degenerates to the Peregrine soliton. In this case, both periods are infinite meaning that the solution is localized both in time and in space.
The vertical line $\alpha_2=1$ corresponds to the Kuznetsov-Ma soliton or a soliton on a finite background. This solution is periodic in $z$ and localised in $t$.
The background becomes zero at the point $\alpha_1=0$ and $\alpha_2=1$. Thus, the soliton on a background turns into an ordinary bright soliton on zero background.
The line $\alpha_1=0$ corresponds to the periodic in $t$ and stationary in $z$ solutions known as $DN$oidal waves. The period is variable and changes along the horizontal line.
The point $\alpha_1=\alpha_2=0$ corresponds to the CW solution. The straight line connecting the points (1/2,1/2) and (0,1) on the plane ($\alpha_1$,$\alpha_2$) corresponds to the family of doubly periodic solutions first presented in \cite{Akh1986} (see Eq.(18) of this work). It was further studied theoretically and experimentally in more recent works \cite{PRA2017,Kim2016}.

For periodic solutions, important physical parameters are the maximal and the minimal amplitudes of the wave profiles. We present here the minimum and the maximum values of $\psi(z,0)$ as a function of distance $z$ at fixed time $t=0$. The $t=0$ is chosen as the point where the wave amplitude changes the most. This can be seen clearly from Fig.\ref{fig1}(a).
In particular, the ratio of these values can serve as a measure of the contrast of amplitude oscillations. From the exact solution, we get:
\begin{align}
\label{psiminA}\psi_{min}=\min_z|\psi(t=0,z)|&=-\sqrt{\alpha_1}+\sqrt{\alpha_2}+\sqrt{\alpha_3},\\
\label{psimaxA}\psi_{max}=\max_z|\psi(t=0,z)|&=\sqrt{\alpha_1}+\sqrt{\alpha_2}+\sqrt{\alpha_3}.
\end{align}
The two surfaces described by Eqs.(\ref{psiminA}-\ref{psimaxA}) are shown in Fig. \ref{fig2}(c). Naturally, they are joined together at the line $\alpha_1=0$ as the $DN$oidal wave does not evolve in time. The minimal amplitude here is $1$. The maximal amplitude of $3$ is reached in the case of the Peregrine solution ($\alpha_1=\alpha_2=1$). This is the expected absolute maximum of the AB solutions at this limit.

For completeness, Fig.\ref{fig2}(d) shows the amplitude of the CW component at $z=0$ extracted from Eq.(\ref{fourierA0}). For the AB solutions, the CW component is always $1$ when \green{$\alpha_3=1$}. 
For all other cases, the CW component is smaller. More generally, when $\alpha_3$ is different from $1$, the estimates can be made in terms of $\alpha_3$. For values of parameters sufficiently far from the cases $\alpha_1=0$ (DN limit) and $\alpha_2=0$ (KM limit), the estimate is $|\hat\psi_0(0)|\approx \sqrt{\alpha_3}$. This means that for the solutions describing the amplification of a small harmonic perturbation over a strong CW, the value of the CW is fixed mainly by the parameter $\alpha_3$.

\subsection{The family of A-type solutions}

As shown in \cite{Akh1987}, A-type solutions, $\alpha_1$ and $\alpha_2$ are complex and it is more convenient to switch to other two parameters $\rho$ and $\eta$ defined as $\alpha_1=\alpha_2^*=\rho+i\eta$. Then the family of solutions depends on three real parameters $\alpha_3>0,\rho,\eta$.
In this case, for the function  $\delta(z)$, we have the following expression:
\begin{equation}
\boxed{\delta(z)=\sqrt{\frac{\alpha_3}{2}(1-\nu)}\sqrt{\frac{1+\dn(\mu z,k)}{1+\nu \cn(\mu z,k)}}\sn(\mu z/2,k)},
\end{equation}
where 
$m=k^2=\displaystyle \frac{1}{2}\left(1-\frac{\eta^2+\rho(\rho-\alpha_3)}{AB}\right)$,
$$A^2=(\alpha_3-\rho)^2+\eta^2,
~~~~ B^2=\rho^2+\eta^2, ~~~~
\nu=\displaystyle\frac{A-B}{A+B},$$
and $\mu=4\sqrt{AB}$.
Clearly, $0<\delta^2<\alpha_3$, by construction.

The phase $\phi(z)$ is given by:
\begin{empheq}[box=\fbox]{align}\label{psiB}
\nonumber\phi(z)=\left(2\rho+\frac{\alpha_3}{\nu}\right)z-&
\frac{\alpha_3}{\nu\mu}\bigg[\Pi(\am(\mu z,k),n,k)- \\
&-\nu\sigma\tan^{-1}\left(\frac{\sd(\mu z,k)}{\sigma} \right) \bigg]
\end{empheq}
where
$n=\displaystyle \frac{\nu^2}{\nu^2-1}$, 
$\sigma=\displaystyle \sqrt{\frac{1-\nu^2}{k^2+(1-k^2)\nu^2}}$, and
$\sd(\mu z,k)=\displaystyle \frac{\sn(\mu z,k)}{\dn(\mu z,k)}$.
Formula (\ref{psiB}) for the phase is the corrected version of the one presented earlier in \cite{Akh1987,Akhbook}.

The function $Q(t,z)$  has the following expression:
\begin{equation}\label{QtypeB}
\boxed{Q(t,z)=sb-c_+\frac{r+\cn(pt,k_q)}{1+r\cn(pt,k_q)}},
\end{equation}
where 
$s=s(z)=\sign\left[\cn(\mu z/2,k)\right]$,
$r=\displaystyle \frac{M-N}{M+N}$,
$$p=\sqrt{MN}=2\displaystyle\sqrt[4]{(\alpha_3-\rho)^2+\eta^2},$$
$$k_q^2=\displaystyle \frac 1 2+2\frac{\rho-\alpha_3}{p^2},
~~~ b=\sqrt{\alpha_3-y},  ~~~  y(z)=\delta^2(z),$$
$$c_\pm=\sqrt{2\left[\sqrt{(y-\rho)^2+\eta^2}\pm(\rho-y)\right]},$$
$$M^2=(2sb+c_+)^2+c_-^2, ~~~ N^2=(2sb-c_+)^2+c_-^2.$$

\begin{figure}[ht]
\centering
\includegraphics[width=\columnwidth]{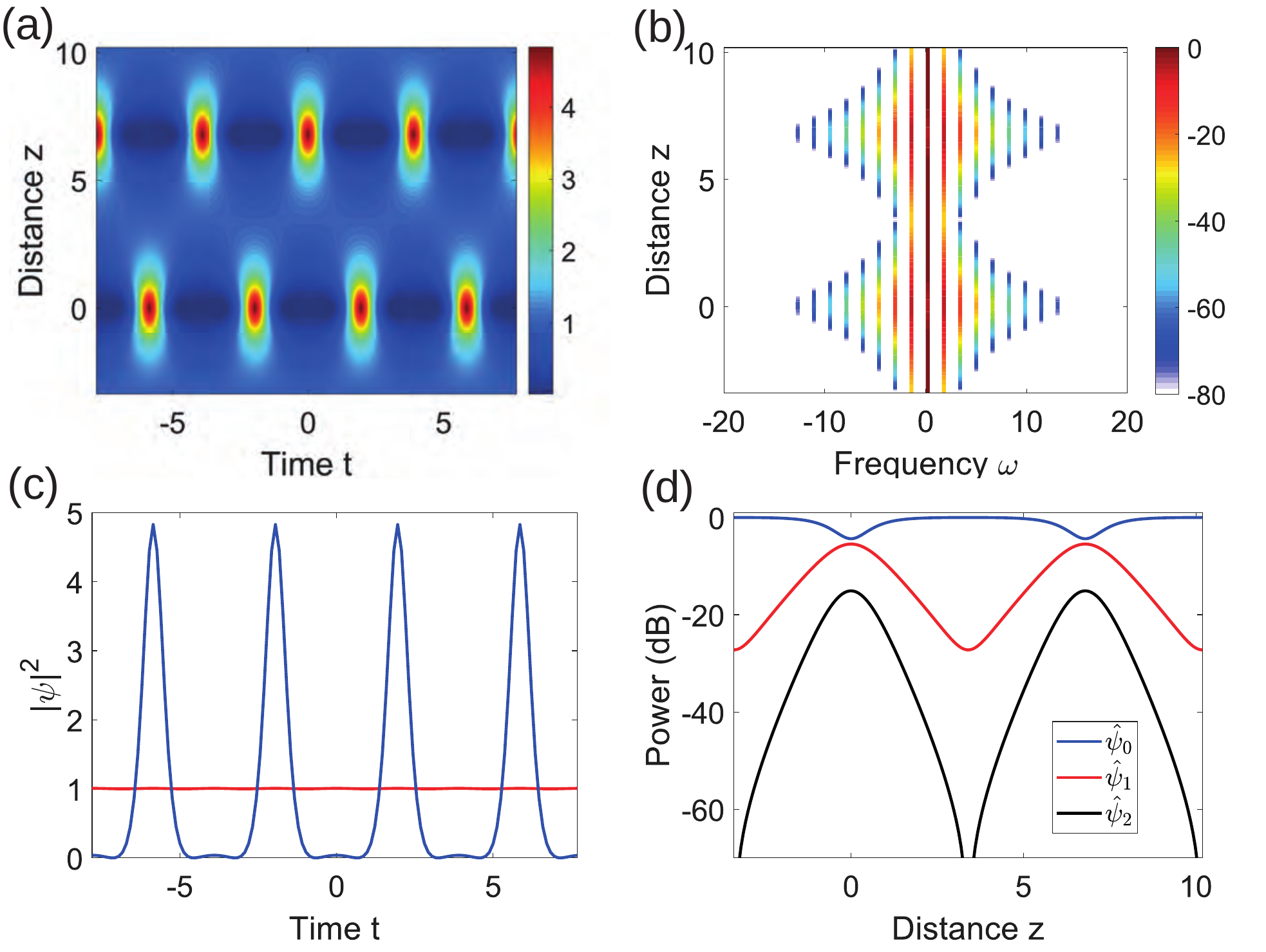}	
\caption{An example of the A-type doubly periodic solution. (a) False color plot of intensity $|\psi(t,z)|^2$. One complete period of evolution is shown. (b) Evolution of the spectrum. (c) The intensity profile, $|\psi(t)|^2$, at $z=-L/4$ (red curve, the minimal modulation) and at $z=0$ (blue curve, the maximal pulse compression). (d) Evolution of the first three Fourier components in logarithmic scale ($20\log_{10}|\hat\psi_k(z)|$). Parameters: $\alpha_3=0.3$, $\rho=0.355$, $\eta=0.073$. The resulting temporal period $T=3.9$ corresponds to the modulation frequency $\omega=2\pi/T=1.61$. The spatial period $L=13.6$.} 
\label{fig3}
\end{figure}

The periods along $z$ ($=L$) and $t$ ($=T$) are:
\begin{equation}
L=\frac{8K(k)}{\mu},\;\;T=\frac{4K(k_q)}{p}.\label{periodB}
\end{equation}
A typical example of the intensity profile evolution of A-type doubly periodic solution is shown in Fig. \ref{fig3}(a). For given set of parameters, this solution also  describes the amplification of a weakly modulated CW ($z=-L/4=-3.4$) towards the generation of a periodic train of pulses ($z=0$). \green{This can be seen by comparing the initial (at $z=-L/4$) and the maximally compressed (at $z=0$) profiles in Fig.\ref{fig3}(c).} However, in contrast to the B-type solution, after the first recurrence back to the initial condition, where the initial CW state is recovered (at $z=L/4=3.4$), the follow up evolution differs qualitatively from the B-type case. Namely, the next growth-decay cycle generates a train of pulses which is shifted by a half of the temporal period in time domain ($z=L/2=6.8$) relative to the pulse train in the first cycle. Snapshots of the initial ($z=-L/4$) field profile and the generated train of pulses  ($z=0$) are shown in Fig. \ref{fig3}(c) by the red and the blue curves, respectively. 

Our analysis provides an elegant analytic description of the FPU recurrence and a symmetry breaking in infinite-dimensional dynamical systems. Here, it takes the form of the transition between the A-type and B-type doubly periodic solutions. Indeed, the spreading of the spectrum from one CW component to several sidebands and the follow up compression of the energy back to the same single component is the manifestation of the FPU recurrence. Switching between the two scenarios of this recurrence while crossing the separatrix is a symmetry breaking. However, we should remember that the transition from one type of orbits to another one occurs in an infinite-dimensional phase space. This transition is far from being a simple copy of symmetry breaking in systems with one degree of freedom. Complexity of these transitions can be seen from some examples given in \cite{PRA2017}.

The freedom of tuning independently the spatial and temporal periods of the solution is a powerful tool in description of variety of physical phenomena. In addition, the third parameter $\alpha_3$ provides the freedom of changing arbitrarily the amplitude of the solutions. Clearly, this family of solutions is more general than the one used in \cite{Kim2016}.

\subsection{Fourier spectra of A-type solutions}
The Fourier coefficients of $Q$ function given by Eq. (\ref{QtypeB}) can be calculated using the technique from \cite{Lang}. These spectra are as follows:
\begin{empheq}[box=\fbox]{align}
\label{fourierB0}\hat Q_0(z)&=sb+\frac{c_+}{r}\left( \frac{\Pi(n,k_q)}{K(k_q)}-1\right),\\
\label{fourierB}\hat Q_\ell(z) &=\frac{c_+\pi \lambda}{2K(k_q)}\cdot 
\begin{cases} \displaystyle s\frac{\sinh(\ell w)}{\sinh(\ell w_0)}, & \mbox{if } \ell \mbox{ is even} \\
 \displaystyle -\frac{\cosh(\ell w)}{\cosh(\ell w_0)}, & \mbox{if } \ell \mbox{ is odd} \end{cases}
\end{empheq}
where
$n=r^2/(r^2-1)$, ~~~$\displaystyle\lambda=\sqrt{\frac{1-r^2}{m_q+r^2(1-m_q)}}$, $$w=\displaystyle\frac{\pi [K(k'_q)-v_0]}{2K(k_q)}, ~~~ w_0=\displaystyle\frac{\pi K(k'_q)}{2K(k_q)},$$ and $v_0=F(\sin^{-1}\sqrt{1-r^2},k'_q)$.

Interestingly, the even and odd Fourier coefficients in (\ref{fourierB}) are different. This is in striking contrast to the case of the B-type solutions.
An example of the evolution of the spectra of A-type solution is shown in Fig. \ref{fig3}(b).  The evolution of the power spectrum is periodic in $z$ with period $L/2$. It shows periodic expansion towards higher order spectral components and recurrence back to the initial spectrum.  The spectra remain symmetric with respect to zero frequency at every $z$. At the point of maximal expansion, a triangular comb of 17 lines can be seen within the range of 80 dB.  
These include the central component $|\hat\psi_0(0)|= 0$ dB, and the first sideband $|\hat\psi_{\pm}1(0)|=-27.0$ dB. The pair of second order sidebands vanishes, $|\hat\psi_{\pm2}(0)|=0=-\infty$ dB. Moreover, all even order sidebands also completely vanish as can be seen from Eq.(\ref{fourierB}). They are recovered at the points of minimal spectral expansion ($z=L/4+\ell L/2$, $\ell=0,\pm1,\ldots$). This can be seen in Fig. \ref{fig3}(d) which shows the evolution of the first three spectral components.

\subsection{Major characteristics of A-type solutions}

The major characteristics of the family of doubly periodic A-type solutions are periods along the $z$ and $t$ axes and the wave amplitudes. 
In this section, we will express them as a function of control parameters of the family $\rho,\eta,\alpha_3$.
As before, we  keep $\alpha_3=1$ fixed. The amplitude can be rescaled when needed using Eq.(\ref{trans}).
When $\alpha_3=1$, the average initial amplitude $|\hat\psi_0(0)|= 1$, as it can be seen from Eq.(\ref{fourierB0}).
Physically, this means that at the point of minimal spectral expansion, the A-type solution describes a weak modulation of a CW of unit amplitude.

\begin{figure}[ht]
	\centering
\includegraphics[width=\columnwidth]{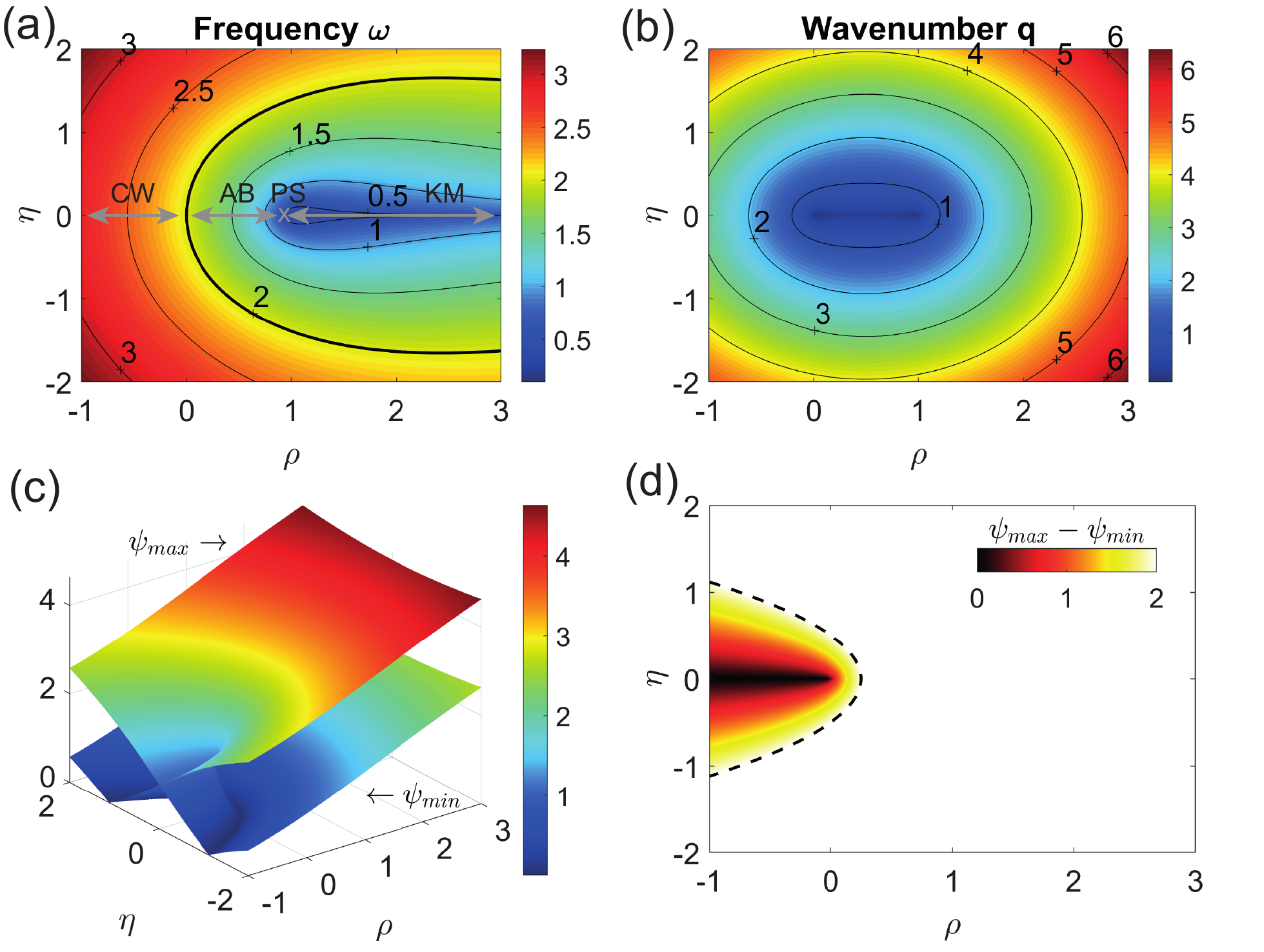}	
	\caption{Major characteristics of A-type doubly periodic solutions vs $\rho$ and $\eta$. False color plots of (a) frequency $\omega=2\pi/T$ and (b) wavenumber $q=2\pi/L$. (c) Maximal ($\psi_{max}$) and minimal ($\psi_{min}$) values of $|\psi(t=0,z)|$. (d) Difference between the maximal and minimal values of $|\psi(t=0,z)|$. Notations in (a): PS - Peregrine solution, CW - continuous wave, AB - Akhmediev breathers, KM - Kuznetsov-Ma solitons, BS - bright soliton.  Here parameter $\alpha_3=1$.}
	\label{fig4}
\end{figure}

Figures \ref{fig4}(a) and \ref{fig4}(b) show the frequency $\omega=2\pi/T$ and the wavenumber $q=2\pi/L$, respectively, calculated from Eq.(\ref{periodB}).
The crucial difference from the B-type case is that there is no frequency cut-off for the A-type solutions.
This means that these solutions can describe the growth of weak perturbations on a constant background even \emph{outside the conventional MI band}, [0,2].
The whole yellow-red area in Fig.\ref{fig4}(a) corresponds to the frequencies above the limiting value $2$.
Comparing Figs.\ref{fig4}(a) and \ref{fig4}(b), we conclude that,  in this region, the higher the frequency $\omega$, the higher is the wavenumber $q$.
Thus, when frequency $\omega$ is well outside the MI band, the solution will oscillate rapidly in $z$ direction. This is a typical feature of non phase-matched four-wave mixing.

Particular cases of this general three-parameter family of solutions can be identified based on spatial and temporal periods. Each of them is still
a family of solutions with lower number (two) of parameters.
When $\eta=0$ and $0<\rho<1$, we get the homoclinic AB solutions. They are periodic in $t$ while period $L$ along $z$-axis tends to infinity.
When $\eta=0$ and $\rho=1$, the solution becomes the Peregrine soliton with no free parameters (except $\alpha_3$). It is localized both in time and in space thus representing a rogue wave. The case $\eta\rightarrow0$ and $\rho>1 $ corresponds to the Kuznetsov-Ma solitons. They are periodic in $z$ and localized in $t$. When $\eta=0$ and $\rho<0 $, the solution becomes the trivial CW  (homogeneous) solution.  Note that if $\eta$ is strictly zero, $\eta=0$,
 then $\rho<1$ because the roots $\alpha_1=\alpha_2=\rho<\alpha_3$ by definition. 

The minimum and the maximum of $\psi(0,z)$ as a function of distance for $t=0$ are given by:
\begin{align}
\label{psiminB} \psi_{min}&=\left|\sqrt{\alpha_3}-\sqrt{2\left[\sqrt{\rho^2+\eta^2}+\rho\right]}\right|,\\
\label{psimaxB} \psi_{max}&=\sqrt{\alpha_3}+\sqrt{2\left[\sqrt{\rho^2+\eta^2}+\rho\right]}.
\end{align}
The two surfaces described by Eqs.(\ref{psiminB}) and (\ref{psimaxB}) are shown in Fig.\ref{fig4}(c). It can be seen from Fig.\ref{fig4}(c) that when $\sqrt{\rho^2+\eta^2}+\rho>\alpha_3/2$, the two surfaces are nearly parallel. Then the difference between the maximal and minimal values saturates to the  maximum value $\psi_{max}-\psi_{max}=2\sqrt{\alpha_3}$. This is shown separately in Fig.\ref{fig4}(d).

\section{Nonlinear stage of linearly stable perturbations}

The fact that the A-type solutions have an arbitrary temporal period, have intriguing effects on the nonlinear stage of MI. As we know well \cite{Benjamin1967,Bespalov1966}, periodic perturbations outside the conventional MI band, $0<\omega<2$, in the linearised problem are not growing. The detailed analysis of A-type solutions leads to a different conclusion, 
as explained below in detail. 

\begin{figure}[ht]
	\centering
		\includegraphics[width=\columnwidth]{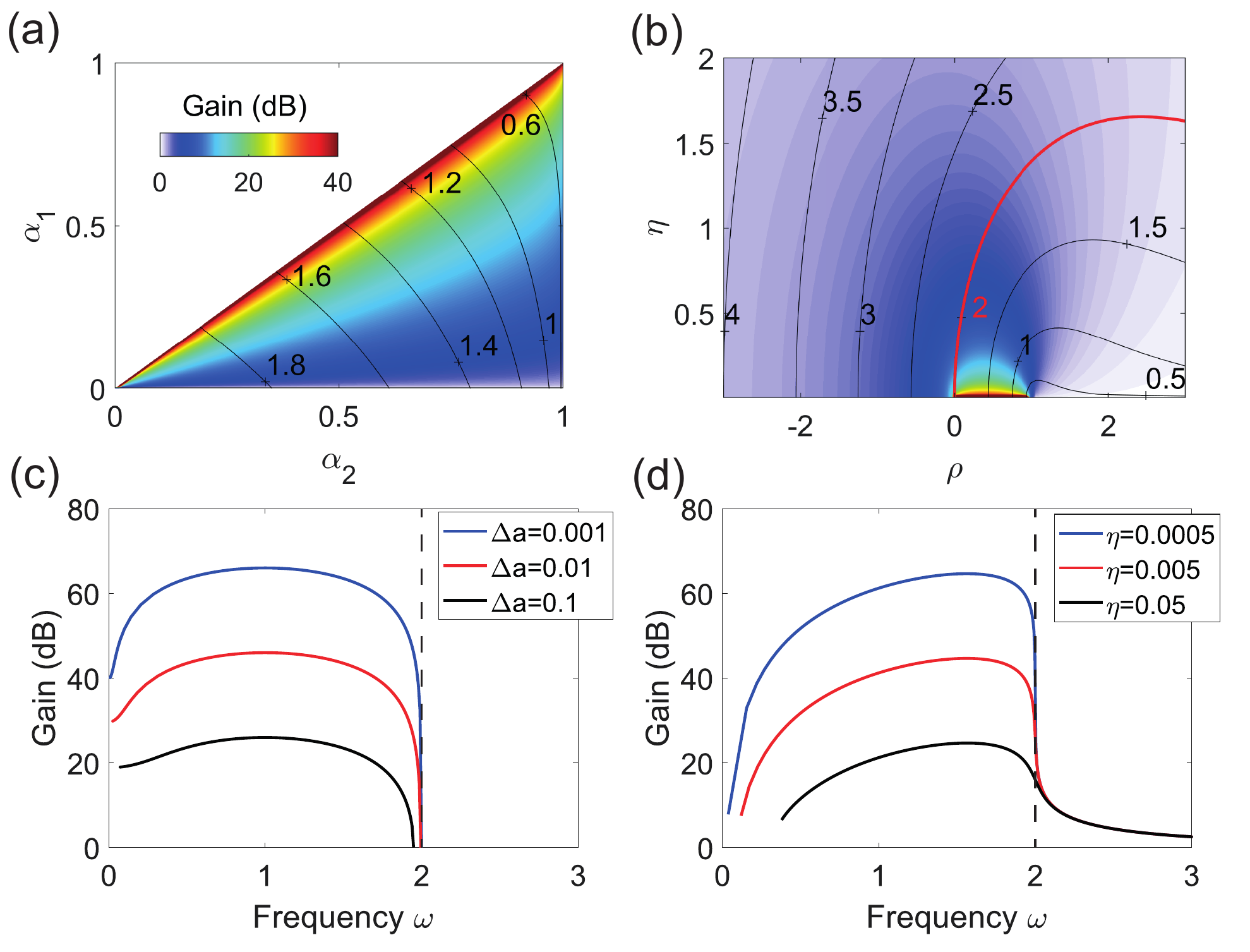}	
\caption{False colour plots of amplification of the first Fourier harmonics $|\hat\psi_1|$ in decibels for (a) the B-type and (b) the A-type doubly periodic solutions. The superimposed level curves are for the constant frequencies $\omega$. Colormap is saturated at 40 dB for readability. (c) Amplification curves on selected straight lines $\alpha_1=\alpha_2 - \Delta a$ in the plot (a). Here $\Delta a<\alpha_2< 1$. 
(d) Amplification curves at selected lines of constant $\eta$ in the plot (b). Here $-3<\rho< 1$. Vertical dashed lines in (c) and (d) show the MI threshold for a CW of unit amplitude. In all cases, $\alpha_3=1$. }
	\label{fig5}
\end{figure}

Figure \ref{fig5} shows the amplification of the first Fourier harmonic for B-type and A-type solutions.
Namely, Fig. \ref{fig5}(a) is a 2D colour coded plot of gain of the first sideband defined as $G=|\hat\psi_1(L/2)/\hat\psi_1(0)|$ in dB for the B-type solution as a function of parameters $(\alpha_1,\alpha_2)$ when $\alpha_3=1$. On the line $\alpha_1=\alpha_2$, the gain defined this way, goes to infinity, because the modulation tends to zero for $z\rightarrow0$ and the period $L\rightarrow\infty$. In order to enhance the readability of the picture, the colour scale is saturated at 40 dB. The lines of constant frequency $\omega$ are superimposed in order to see their value relative to the MI band. The gain is decreasing away from the AB limit, and completely vanishes on the line $\alpha_1=0$. The solution at this line describes a $z$-stationary solution. Three gain curves along the lines $\alpha_1=\alpha_2 - \Delta a$  parallel to the line of AB solutions are shown in Fig. \ref{fig5}(c). The highest amplification is observed on the line that is closest to the AB limit. As these curves show, all frequencies within the MI band are amplified as we would expect from the standard MI analysis. Gain is zero outside of this band also in agreement with the standard MI theory.

The situation is different for the A-type solutions. Figure \ref{fig5}(c) shows the colour coded 2D plot of gain defined the same way $G=|\hat\psi_1(0)/\hat\psi_1(-L/4)|$ in dB for A-type solution as a function of parameters $(\rho,\eta)$. The third parameter $\alpha_3=1$. The lines of constant frequency $\omega$ are superimposed on this plot allowing us to identify the frequencies inside and outside the MI band. The red curve in this plot corresponds to the upper limit of conventional MI band, $\omega=2$. As in the case of the B-type solutions, the gain tends to infinity within the conventional MI band  ($0<\omega<2$) when $\eta\rightarrow 0$ and $0<\rho<1$. The $z$-period on this line is also infinite as it should be for the AB solutions.   The colour scale here is the same as in Fig. \ref{fig5}(a). It is saturated at 40 dB. The value of gain decreases away from this area, as expected. However, it never vanishes to zero which is an important conclusion of our analysis. Indeed, this is the most striking feature of A-type solutions. They describe \emph{amplification and parametric gain outside the standard MI band}. This feature is further evident from Fig. \ref{fig5}(d), showing gain curves at small constant values of $\eta$. As we can see from these curves, there is no cut-off at $\omega=2$, but rather a smooth transition to a region of small gain for $\omega>2$. This gain slowly drops at large frequencies, being nearly independent on $\eta$.

\begin{figure}[!t]
	\centering
		\includegraphics[width=\columnwidth]{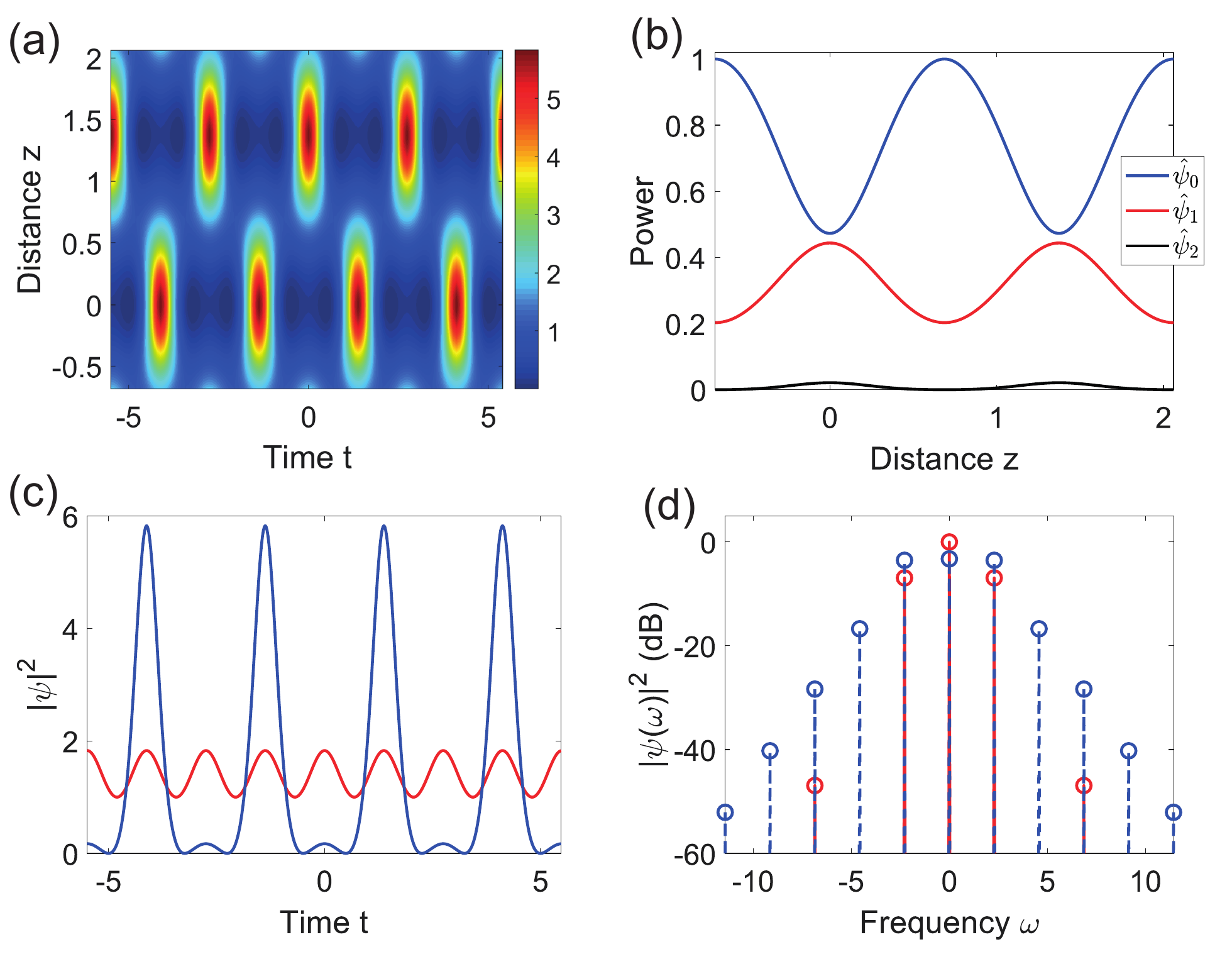}
	\caption{Evolution of the A-type doubly periodic solution illustrating the amplification outside the conventional MI band. (a) False color plot of intensity $|\psi(z,t)|^2$. (b) Evolution of the first three Fourier components. (c) Intensity profiles $|\psi(t)|^2$ at $z=-L/4$ (red curve, minimum of modulation) and at $z=0$ (blue curve, maximum pulse compression). (d) Input (red) and output (blue) spectra. The input spectrum has a pair of small second sidebands. Parameters of the solution: $\rho=0$, $\eta=1$, $\alpha_3=1$. In this case, the temporal period $T=2.74$ (modulation frequency $\omega=2.287>2$) and the spatial period $L=2.75$.}
\label{fig6}
\end{figure} 

An example of the A-type solution illustrating the phenomenon of the first sideband amplification outside the instability band and the follow up recurrence is shown in Fig. \ref{fig6}.
The evolution starts with CW field with small but finite nearly sinusoidal perturbation. This initial condition is shown by red curve in Fig. \ref{fig6}(c). The amplitude of modulation grows transforming the wave profile to a train of pulses. The maximum amplitude of pulses is reached after a quarter of the $z$-period at the point $z=0$. The transformed profile is shown as blue curve in Fig. \ref{fig6}(c). Periodicity takes the profile back to the initial shape after a half of the $z$-period. This occurs at $z=L/4$. However, the phase of the sinusoid is now shifted in time. The next pulse compression point is $z=L/2$. The maxima of the pulses are now located at the position of minima of the previously compressed profile although the shape remains the same.   

The evolution of the lowest order Fourier components during this process is shown in Fig. \ref{fig6}(b). Amplification of the first sideband in the first quarter of the period is clearly seen. Its power at the point of maximum increases by $g=|\hat\psi_1(0)/\hat\psi_1(-L/4)|^2=2.19=3.4$ dB. This happens at the expense of significantly depleted CW component. The process is periodic thus leading to periodic evolution of spectral components. Figure \ref{fig6}(d) shows the spectral content of the initial condition (red circles) and the maximally compressed pulse train profile (blue circles). A remarkable observation that may lead to important applications is that even harmonics are zero at the input field. This is clearly seen in Fig.\ref{fig6}(d). They are generated in evolution and become the strongest at the point of maximal compression still keeping the total spectrum within the triangular shape (blue points). In the time domain, this spectral feature corresponds to the period-doubling of the field intensity. This can be seen clearly in Fig. \ref{fig6}(c). Period of the blue curve is twice the period of the red curve. This is different from the case shown in Fig.\ref{fig1}(c) where periods of the initial condition and the resulting train of pulses coincide. Clearly, the frequency is halved when period doubling takes place. This effect is opposite to the frequency doubling phenomenon considered earlier in \cite{OC2010} once again demonstrating the richness of phenomena contained in the family of doubly periodic solutions.

\begin{figure}[!t]
	\centering
\includegraphics[width=\columnwidth]{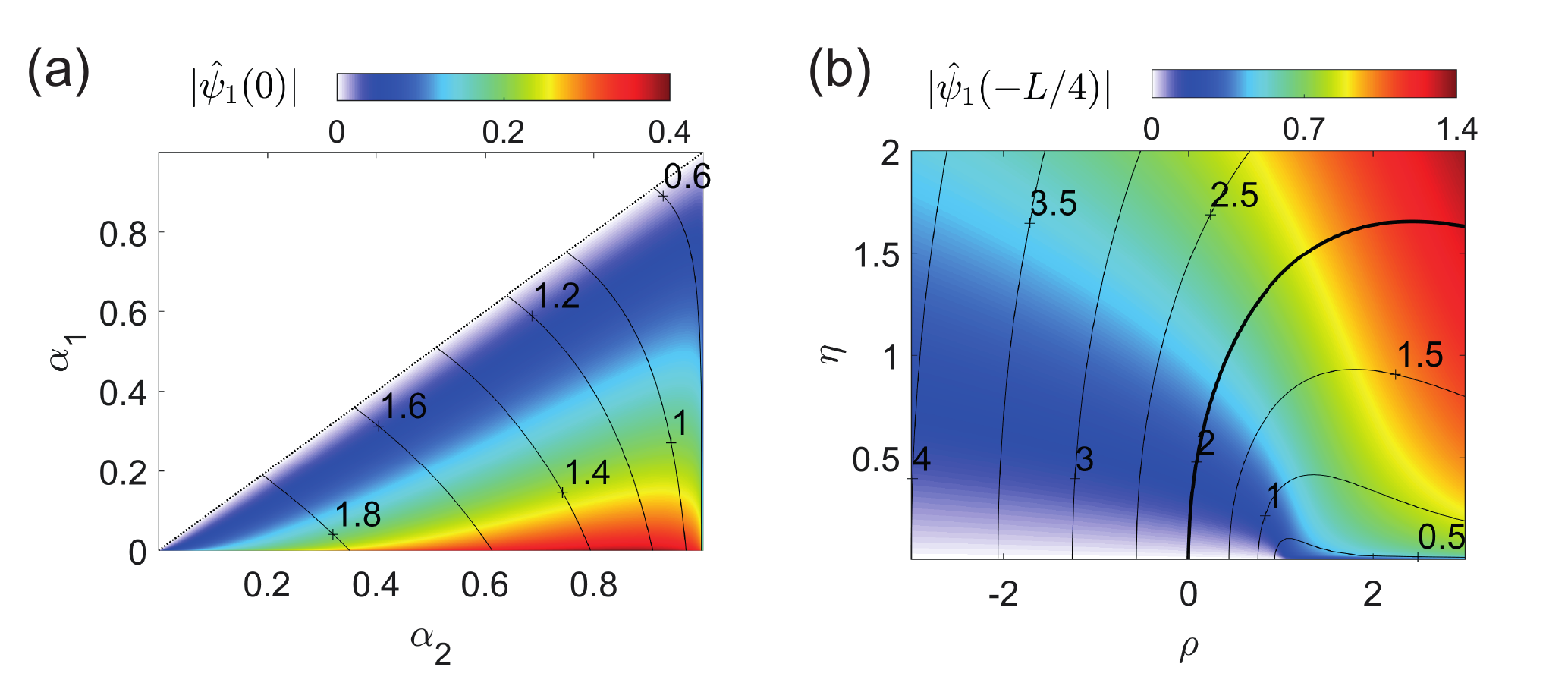}
\caption{
	False colour plots of minimum amplitude of the first Fourier harmonics $|\hat\psi_1|$ in the plane of free parameters of the family of solutions: (a) B-type solutions (minimum sidebands obtained at $z=0$; parameters $\alpha_1, \alpha_2$); (b) A-type solutions (minimum sidebands obtained at $z=-L/4$; parameters $\rho, \eta$). Here $\alpha_3 =1$.}
	
\label{fig7}
\end{figure}

The example shown in Fig. \ref{fig6}(b) corresponds to a case where the initial amplitude of the first-order sideband is relatively large. One might wonder, how small the sideband could be to still achieve amplification out-of-instability-band. In order to answer this question and further reveal the difference between the behaviors of A-type and B-type solutions concerning the out-band amplification, we show, in Fig.\ref{fig7}, the minimum value over $z$ of the first-order Fourier harmonic $|\hat\psi_1|$. This corresponds indeed to the smallest sideband that can be amplified (along with the harmonics that are involved in the solutions) for any given choice of the parameters $\alpha_1, \alpha_2$ for B-type, or $\rho, \eta$ for A-type solutions, assuming $\alpha_3=1$ as before. Figure \ref{fig7}(a), which is related to type B solutions, shows that the sideband amplitude is defined only inside the conventional MI band $\omega<2$, ranging from arbitrarily small values (close to AB, $\alpha_1 \simeq \alpha_2$) to relatively large values when the solution is a deformation of DN-oidal solution ($\alpha_1=0$, see Fig. \ref{fig2}) with large $\alpha_2$. As shown in Fig. \ref{fig7}(b), for the A-type solutions, small sideband amplitudes are obtained close to the horizontal axis ($\eta \simeq 0$) for $\rho<1$ where, at variance with previous case, they exist across the band edge limit (parametric curve $\omega=2$) near the origin of the plane $(\rho,\eta)$. As a consequence, one can still amplify even small sidebands, although it must be considered that the gain abruptly drops across such threshold (see Fig. \ref{fig5}(b) and the blue curve in Fig. \ref{fig5}(d)), whereas the period $L$ decreases. This regime is similar to a parametric amplification characterized by large phase mismatch. Larger values of $\eta$ correspond, in the out-band region, to a rapid increase of the sideband to values which are typically larger than those for B-type solution, and can become comparable to the amplitude of the CW. This is the regime of amplification of large modulations, which, by definition, cannot be described by the standard MI linear stability analysis.

Finally, we give an important argument that helps to get insight into the fact that only A-type solutions exhibit gain outside the conventional MI gain bandwidth.
To this end, we resort to the interpretation of MI as a symmetry breaking process, where the broken symmetry appears as the parameter $\omega$ is decreased below the threshold $\omega=2$ \cite{Mus2018}. As also discussed above, the phase with broken symmetry ($\omega<2$) is characterized by the coexistence of A-type and B-type evolutions \cite{Mus2018}. Conversely, for $\omega>2$, only one type of evolutions exist, which can be considered, for $\omega \gg 2$, a small deformation of {\em strictly linear} solutions. The latter can be easily obtained from the NLSE (\ref{nlse}) by dropping the nonlinear term $|\psi|^2\psi$. By considering, for the sake of simplicity, the initial condition $\psi(t,0)=\psi_0 + c_1 \exp(i \omega t) + c_{1} \exp(-i \omega t)$ containing the pump $\psi_0$ and a single symmetric pair of sidebands with complex amplitude $c_1$, 
the linear solution reads as
\begin{equation} \label{lin}
\psi(z,t)= \psi_0 + \left[ c_1 \exp(i \omega t) + c_{1} \exp(-i \omega t) \right] \exp \left(-i \frac{\omega^2}{2} z\right).
\end{equation}
Equation (\ref{lin}) shows that the dispersion is responsible for a continuous phase rotation of the modulation, which goes through alternating states of amplitude and frequency modulation, characterised by relative phase between the sideband and the pump of $0,\pi$ and $\pm \pi/2$, respectively \cite{TW1991b}. In particular, two successive states of amplitude modulation, obtained at relative distance $\Delta z=2\pi/\omega^2$ (half spatial period of solution in Eq. (\ref{lin})) such that they are phase shifted by $\omega^2 \Delta z/2=\pi$, exhibit a temporal shift of half period in the intensity pattern, similar to the case of A-type solutions. From this point of view, nonlinear solutions of A-type can be considered as the nonlinear dressing of linear solutions, which exist for $\omega>2$ and are smoothly continued into the phase with broken symmetry ($\omega<2$). Conversely, the B-type solutions have genuine nonlinear origin, bearing no analogy to linear solutions. Instead, they appear only in conjunction with the onset of MI, due to the symmetry breaking nature of the phenomenon.


\section{Conclusion}

We studied, in detail, the three parameter family of doubly periodic solutions of the nonlinear Schr\"odinger equation, originally derived by Akhmediev, Eleonskii and Kulagin \cite{Akh1987}.  We reveal the richness of physical phenomena that is contained within this family. The three free parameters of this family allow us to control arbitrarily the spatial and temporal periods of the family and the amplitude of the resulting periodic profiles.

We discovered several new physical phenomena within this family. These include modulation instability outside of the standard instability band known from the classical works of Benjamin-Feir \cite{Benjamin1967} and Bespalov-Talanov \cite{Bespalov1966}. Using this expanded knowledge of modulation instability applied to the classical NLSE will lead to better understanding of nonlinear phenomena and their new applications both in optics and water waves.

One of the major advances of our present work is calculation of Fourier components of periodic field in explicit form. These analytic expressions will lead to further progress in physical applications of the NLSE related phenomena in fibre optics and water waves.

From theoretical point of view, our analysis provides more physical understanding of fields generated by periodic initial conditions. Previous approaches based on general solution expressed in terms of theta functions did not lead to a significant progress as it is difficult to extract physically relevant results from general solutions. It is a better idea to construct general periodic solutions from fundamental ones in the way similar to construction of multi-soliton solutions from its fundamental constituents - single solitons. There are several techniques that can be used for this aim such as Darboux or Backlund transformations. Being equipped with the family of the fundamental doubly periodic solutions of the NLSE, it will be possible to construct more complicated solutions using this one as a starting point. 

\section*{ACKNOWLEDGMENTS}
This work was partly supported by IRCICA, by the ``Agence Nationale de la Recherche" through the LABEX ``Centre Europeen pour les Mathematiques, la Physique et leurs Interactions" (CEMPI) and EQUIPEX ``Fibres optiques pour les hauts flux" (FLUX) through the ``Programme Investissements d'Avenir", by the Ministry of Higher Education and Research, ``Hauts de France" council and European Regional Development Fund (ERDF) through the ``Contrat de Projets Etat-Region" (CPER Photonics for Society, P4S) and the HEAFISY project.

\end{document}